\documentstyle[11pt,newpasp,twoside,epsf]{article}
\def\edcomment#1{\iffalse\marginpar{\raggedright\sl#1\/}\else\relax\fi}
\reversemarginpar

\begin{document}
\title{Numerical Convergence of  
  Hydrodynamical SPH Simulations of Cooling Clusters}
\author{Riccardo Valdarnini}
\affil{ SISSA Via Beirut 2-4 34014, Trieste, Italy}

\begin{abstract}
The results from hydrodynamical TREESPH simulations of galaxy clusters are used
to investigate the dependence of the final cluster X-ray properties 
upon the numerical resolution and the assumed star formation models for the 
cooled gas. When cold gas particles are allowed to convert into stars the final
 gas profiles show a well defined core radius and the temperature profiles are
 nearly flat. 
A comparison between runs with different star formation methods 
shows that the results of simulations,
based on star formation methods in which gas conversion into stars is
controlled by an efficiency parameter $c_{\star}$, are sensitive to the
simulation numerical resolution.
In this respect star formation methods based instead on a local density
threshold, as in Navarro \& White (1993), are shown to give more stable 
results. Final X-ray luminosities are found to be numerically stable,
with uncertainties of a factor $\sim 2$.
\end{abstract}

\section{Introduction}

Galaxy clusters are the largest virialized structures known in the universe
and also bright X-ray sources.
Useful cosmological information can be obtained from the statistical 
properties of the ensemble of X-ray clusters.
X-ray observations of cluster number counts, the X-ray temperature function
(Henry \& Arnoud 1991; Edge et al. 1990; Henry 1997)
 and the X-ray luminosity function 
(Rosati et al. 1998; Ebeling et al. 1998)
 are powerful probes to constrain the values of
the cosmological parameters $\Omega_0$ and $\sigma_8$ 
(Henry \& Arnoud 1991; White, Efstathiou, \& Frenk 1993; Bahcall \& Fan 1998;
Eke, Cole, \& Frenk 1996; Kitayama, Sasaki, \& Suto 1998)

Hydrodynamical simulations have been widely used to predict for different
theoretical frameworks the time evolution of the gas and temperature 
distributions. 
 The numerical methods used are either 
Eulerian  (Cen \& Ostriker 1994; Anninos \& Norman 1996; Bryan \& Norman  1998;
Kang et al. 1994; Bryan et al. 1994, Cen 1997; Cen et al. 1995)
with a fixed or adaptative grid, or Lagrangian 
(Evrard 1988; 1990; Thomas \& Couchman 1992; Navarro, Frenk, \& White 1995; 
Eke, Navarro, \& Frenk 1998; Katz \& White 1993; Yoshikawa, Itoh, \& Suto 1998;
Valdarnini, Ghizzardi, \& Bonometto 1999)
In these simulations the gas component is treated as a single adiabatic fluid,
without taking into account the effects of radiative cooling, and the physical
processes which can be modelled are merging, substructure formation, shocks
and compressional heating of the gas. A comparison between different 
numerical simulations shows that they are successful in reproducing 
the gross features of the cluster properties (Frenk et al. 1999).
With increasing availability of computational power, numerical hydrodynamical
simulations have been attempted to model the effects of radiative 
cooling on the gas in the formation and evolution of cluster galaxies
(Katz \& White 1993; Suginohara \& Ostriker 1998; Anninos \& Norman 1996; 
Yoshikawa, Jing, \& Suto 2000; Pearce et al. 2000; Lewis et al. 2000)
The numerical problems posed by the inclusion of 
gas cooling are challenging, mainly because the required increased spatial
resolution also requires that one keeps under control two-body heating 
mechanisms.
Previous simulations have produced some conflicting results 
(Yoshikawa, Jing, \& Suto 2000; Pearce et al. 2000),
 thus the question of the minimum resolution in this
kind of simulations is still to be fully settled.

In this contribution I will show the preliminary results that have been 
obtained from a series of 
hydrodynamical SPH simulations of galaxy clusters. The simulations  have 
different numerical resolution and
include the effects on the gas component of radiative losses, star formation 
and energy feedback from SN. 
 Final profiles are compared in order to assess the 
effects of
numerical resolution, or different star formation prescriptions, on the cluster
X-ray variables.

\section{Simulations}
In a previous paper (Valdarnini et al. 1999, hereafter VGB) a 
large set of hydrodynamical simulations was 
used to study the global X-ray cluster morphology and its evolution.
The simulations were run using a TREESPH code with no gas cooling or heating.
 In order to assess the numerical reliability of the numerical integrations,
four different clusters were selected as a representative sample of all the 
simulation clusters.
For this cluster sample a large set of different integrations was performed 
by varying two numerical input parameters: the number of particles and the 
softening parameters. 
I refer to VGB for a detailed description of the simulations.
In Valdarnini (2001) I have used the same cluster sample to study the effects of
 including in the simulations additional physics such as gas cooling and star
 formation.
I will report here the results for the cluster with label $\Lambda$CDM$00$ in 
VGB. This is the most massive cluster ($M_v \simeq 1.5 \cdot 10^{15} 
M_{\odot}$) extracted from a cosmological $\Lambda$CDM N-body 
simulation with size $L=200h^{-1}Mpc$, matter density    
 $\Omega_m=0.3$ and Hubble constant $H_0=70  Km~ sec^{-1}~ Mpc^{-1}$.
In order to check the effects of radiative cooling for this
cluster a set of TREESPH simulations was performed,
 with initial  conditions provided by the cosmological simulation,
 and different values of the gas softening parameter $\varepsilon_g$ 
and number of gas particles $N_g$. 
For this cluster Table 1 reports the values of $\varepsilon_g$ and $N_g$ 
for the simulation runs. The 
generic simulation
has cluster index cl$00-j$, with $j=00,01,...,05$. The cluster cl$00-00$ is
the reference case without cooling. 
The effects of radiative cooling are modelled in these simulations by 
adding to the SPH thermal energy equation an energy-sink term.  
 The total cooling function includes 
contribution from
 recombination and collisional excitation, bremsstrahlung and inverse
Compton cooling. 

Allowing the gas to cool radiatively will produce dense clumps of gas
at low temperatures ($\simeq 10^4 \kern 2pt\hbox{}^\circ{\kern -2pt K} $). 
 In these regions the gas
will be thermally unstable and will likely meet the physical 
conditions to form stars.
In TREESPH simulations star formation (SF) processes have been implemented 
using different approaches. 
According to Katz, Weinberg, \& Hernquist (1996, hereafter KWH) 
 a gas particle is in a star forming region if the flow
is convergent and the local sound crossing time is larger 
than the dynamical time (i.e. the fluid is Jeans unstable). 
  In a simplified version  
Navarro \& White (1993 , hereafter NW) assume as a sufficient condition 
 that a  gas particle must be in a convergent flow and its density 
exceeds a threshold  ($\rho_g >\rho_{g,c}=7 \cdot 10^{-26} gr cm ^{-3}$).

\begin{figure}
\begin{center}
\plotone{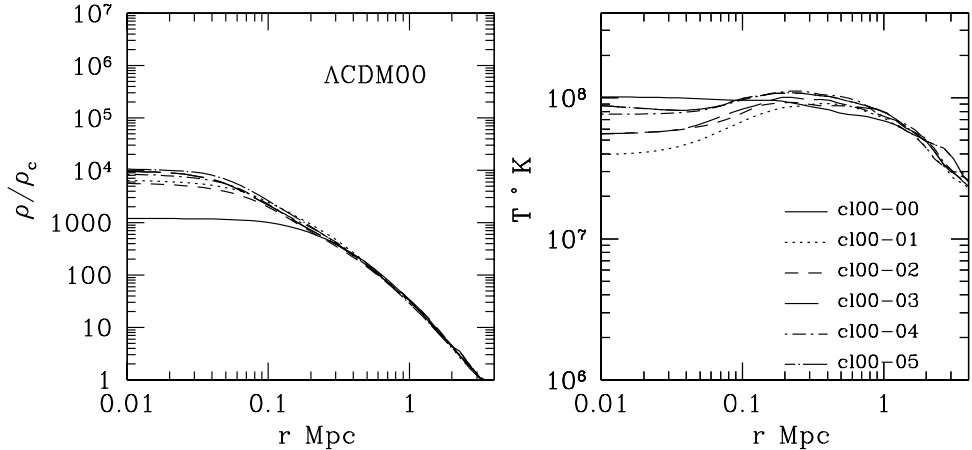} 
\end{center}
\caption{
 Final density and temperature profiles in the simulation 
runs including radiative cooling and star formation.  
  Density is in units of the critical density.
 Different curves are for integrations with different numbers of  
 particles  and different softening parameters (see Table 1).
The simulation run with index $-00$ is the integration without cooling.}
\end{figure}

If a gas particle meets these criteria then it is selected as an eligible 
particle to form stars.
The local star formation rate (SFR) obeys the equation 
 \begin{equation} 
 d\rho_{g}/dt=-c_{\star} \rho_g /\tau_g= -d \rho_{\star} /dt~,
\label{eq:rg}
\end{equation}
where $\rho_g$ is the gas density, $\rho_{\star}$ is the  
star density , $c_{\star}$ is a characteristic dimensionless efficiency 
parameter, $\tau_g$ is the local collapse time and it is the maximum 
of the local cooling time $\tau_c$ and the dynamical time $\tau_d$. 
Gas particles 
with $ T {\lower.5ex\hbox{$\; \buildrel < \over \sim \;$}}
10^4 \kern 2pt\hbox{}^\circ{\kern -2pt K} $ 
 have long cooling times and $\tau_g=\tau_d$.
At each step the probability that a gas particle will form stars in a time 
step $\Delta t$ is compared with a uniform random number. 
 If the test is successfull  then a mass fraction
$\varepsilon_{\star}$ of the gas mass is converted into a new 
collisionless particle. This star particle has the position, velocity and
gravitational softening of the original gas particle. 
Typical assumed values are  $\varepsilon_{\star}=1/3 $ and $c_{\star}=0.1$ (KWH).
For the local SFR, NW adopt Eq. 1 with $c_{\star}=1$, $\tau_g=
\tau_d$ and $\varepsilon_{\star}=1/2$ when a gas particle can convert part of its mass into 
a star particle. 

The numerical tests have been performed 
following the  NW prescriptions for selecting gas particles 
which can form stars.
Once a star particle is created it can release energy into the
 surrounding gas through supernova (SN) explosions. 
 All the stars with mass above $ 8 M_{\odot}$ end as a SN, leaving a 
$1.4 M_{\odot}$ remnant.
The SN energy ($\simeq 10^{51} erg$) is released gradually into the gas 
according to the lifetime of stars of different masses.

\begin{figure}
\begin{center}
\plotone{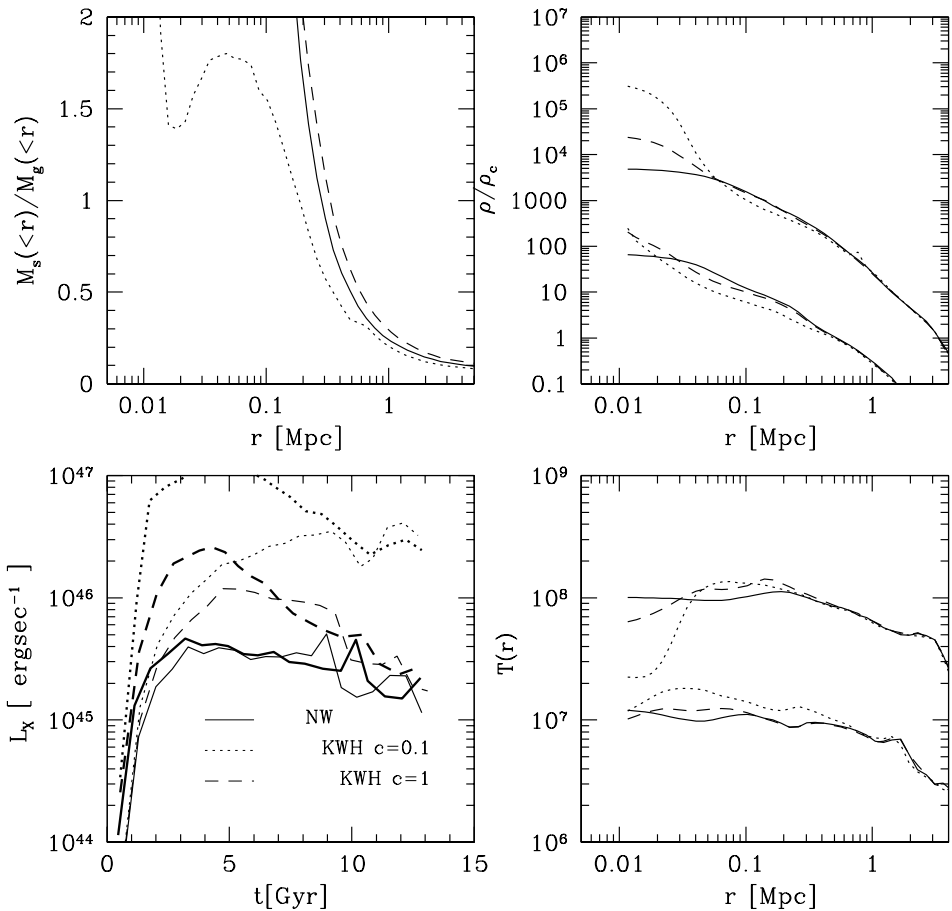}
\end{center}
\caption{
Plots showing several cluster properties in simulation runs
 with different SF prescriptions. The simulation parameters are those of
cl$00-05$ (see Table 1).
The continuous line refers to the NW method, the others to KWH with
different $c_{\star}$ ( $c$ in the bottom left panel).
{\it Top left}: ratio of the star mass within the radius $r$ over the
gas mass within $r$.
 {\it Top right}: final radial density behavior for the gas component.
The simulation results are compared with the
corresponding high-resolution runs (H).
The H simulations  have the same SF parameters of the parent
simulations, but the numerical parameters are given in the
last row of Table 1.
To facilitate a comparison
the radial profiles of the high resolution results
 have been shifted
downward by $10^k$, $k=2$ for densities and $k=1$ for temperatures.
 {\it Bottom right}: Radial temperature profiles.
{\it  Bottom left }: X-ray luminosity versus time, the thick lines
correspond to the high resolution simulations.}
\end{figure}

\section{Results}
The final radial gas density and temperature profiles are 
shown in Fig. 1. 
As can be inferred from  Table 1, the numerical strategy has been to perform
 runs of increasing spatial resolution in order to resolve the core
radius of the gas density profile. 
The most important result is that the inclusion of a 
star formation prescription has been effective in removing the 
cold gas particles 
($ T {\lower.5ex\hbox{$\; \buildrel < \over \sim \;$}}
10^4 \kern 2pt\hbox{}^\circ{\kern -2pt K} $) 
from the cluster center.
In all the simulations the gas density profiles have a well-defined 
core radius, with size $r_c \simeq 50-100 Kpc$, approximately $0.05$ 
of the virial radius $r_v$ ($\simeq 2.9 Mpc$). 
The shapes of the temperature profiles
show that convergence is achieved for $N_g 
{\lower.5ex\hbox{$\; \buildrel > \over \sim \;$}} 20,000$.
All the central values for the gas temperatures at $r=10Kpc$ are within a 
factor $\simeq 1.5$ for the highest resolution simulation runs.
The profiles increase inwards from the virial radius up to 
$ \simeq 100-200 Kpc$. 
Thereafter the profiles stay almost flat, or with a modest decline 
in $T(r)$ towards $r=0$. 
There is not a strong drop of the temperature in the very
central region.
Final X-ray luminosities are quite stable 
($L_X \simeq 2 \cdot 10^{45} erg sec^{-1}$) versus the simulation numerical 
resolution.  Compared to the non-radiative run the luminosities 
 increase by a factor $\simeq 2$. 

A different question 
is the sensitivity of the estimated cluster properties to the 
numerical method used to describe star formation  
in the hydrodynamical simulations.
To this end three simulations were performed 
with the same numerical parameters as for the cl$00-05$ run, but using
different SF methods or parameters. 
The NW simulation (I) is the standard case  showed in Fig. 1.
In the other two runs (II and III) the KWH prescription is adopted for
converting gas particles into stars, but with 
 different values of the star formation efficiency parameter
$c_{\star}$:~$0.1$ and $1.$ (see Fig. 2).
The most important differences are found between the KWH  simulations
with different $c_{\star}$. The differences are
dramatic in the final X-ray luminosities, which differ by a factor 
$ \simeq 40$.
The source of this discrepancy relies in the different gas density profiles,
which have substantial differences in the cluster core regions for
$ r {\lower.5ex\hbox{$\; \buildrel < \over \sim \;$}} 100 Kpc$. 
These differences are localized at the cluster center, 
beyond $ r \simeq 100 Kpc$ all the profiles converge, as it is 
shown in the plots of Fig. 2.
The temperature profiles have a peak value of $ \simeq 10^8 
\kern 2pt\hbox{}^\circ{\kern -2pt K}$ at 
$ r \simeq 100Kpc$ and thereafter decline outward by a factor $\sim 2$ 
out to $r_v$. Below $\sim 100Kpc $ the profiles instead show  
large differences. Compared to the NW run the simulation 
with $c_{\star}=0.1$ has gas temperatures which decline inwards by a 
factor $\sim 10$ from $\sim 100Kpc$ down to $r\sim 10Kpc$.
These radial decays follow because of the less efficient conversion
of the cooled gas into stars compared to the NW run.
There is a remarkable agreement for the ratio of cluster mass locked into
 stars to the gas mass, which is $ \simeq 10 \%$  at $r_v$ for all the 
runs considered.
In order to assess the effects of numerical resolution upon 
final results  simulations I, II and III have been run again   
but with a number of particles increased by a factor $ \simeq 3$ (cl$00-05$H).
 These simulations will be referred as IH, IIH and IIIH, respectively. 
The simulation results are shown in Fig.~2. For simulations
IH there are not appreciable differences in the radial profiles.
This confirms the previous results, that is the NW runs have reached 
numerical convergence in the physical variables for the numerical
parameters of the cl$00-05$ simulation of Table 1. The profiles of 
simulation IIH are instead different from those of run II at 
$r {\lower.5ex\hbox{$\; \buildrel < \over \sim \;$}}
50 Kpc$. The strong drop in $T(r)$ has been removed and 
the gas density profile is much closer to the NW one. 
Simulation IIIH yields final profiles very similar to the ones of 
the parent simulation.
The bottom left panel of Fig.~2 shows that high resolution
runs have final X-ray luminosities which can differ within a factor $\sim 2$
from the parent simulations.

 To summarize, the above results demonstrate that simulations I and III 
 have an adequate numerical resolution to reliably predict 
X-ray cluster properties, such as the X-ray luminosity.
For simulation II ( KWH  with $c_{\star}=0.1$ ) there are large differences 
at the cluster core between the final profiles when the numerical 
resolution is increased. 
For the simulation runs I and III $L_X \simeq 10^{45} erg sec^{-1}$,
while $L_X \simeq 4 \cdot 10^{46} erg sec^{-1}$ for the run with $c_{\star}=0.1$.
Thus simulations I and III are consistent at the 
$1\sigma$  level with the $ L_{bX}-T_X$ relation derived from a sample of 
cooling flow clusters (Allen \& Fabian 1998). For simulation II  
 the predicted cluster temperature is outside the $2\sigma$ limits.

\begin{table}
\begin{tabular}{ccccccccc}
$\Lambda$CDM00 &  $\varepsilon_g^{~a}$ & $m_g^{~b}$ & $m_d^{~c}$ & $N_g^{~d}$ & 
$N_d^{~e}$ & $N_T^{~f}$ & $\theta^g$ & $z_{in}^{~h}$ \\ 
\hline
cl00-00 & 56 & $3.01\cdot10^{10}$ & $2.64\cdot10^{11}$ & 5503 &6295&16463&
0.7& 10.\\ 
cl00-01 & 56 & $3.01\cdot10^{10}$ & $2.64\cdot10^{11}$ & 5503 &6295&16463&
0.7& 10.\\ 
cl00-02 & 28 & $3.01\cdot10^{10}$ & $2.64\cdot10^{11}$ & 5503 &6295&16463&
0.7& 10.\\ 
cl00-03 & 21 &$1.45\cdot10^{10}$ & $1.28\cdot10^{11}$ & 11480&14208&35408&
0.7& 10.\\ 
cl00-04 &15.4&$1.45\cdot10^{10}$ & $1.28\cdot10^{11}$ & 11480&14208&35408&
0.7& 10.\\ 
cl00-05&21&$7.47\cdot10^{9}$ & $6.57\cdot10^{10}$ & 22575&25391&67388&
0.7& 19.\\ 
cl00-05H&10.5&$2.45\cdot10^9$& $2.12\cdot10^{10}$& 69599&74983&204799&$1.0$ 
& 29.\\
\end{tabular}

\caption{Simulation parameters of the test runs for
the $\Lambda$CDM00 cluster. cl$00-00$ is the reference case with no
cooling, taken from VGB. $^{a}$: gravitational softening parameter for the 
gas in $h^{-1}$~Kpc.  $^{b}$: mass of the gas particles in $h^{-1} M_{\odot}$ 
(the cosmology is for $\Omega_m=0.3$ and $h=0.7$).
$^c$ : mass of the dark particles. $^d$: number of gas particles. 
 $^e$ : as in the previous column but for 
dark particles. $^f$: total number of simulation particles. 
$^g$: value of the treecode gravitational tolerance parameter. 
$^h$ : initial redshift for the simulation.  The last row gives the numerical 
parameters for the high-resolution runs used to test different SF methods. 
For these runs gravitational quadrupole corrections were enabled.  
}  
\end{table}

\end{document}